\documentclass[aps,pra,twocolumn,noshowpacs,showkeys]{revtex4-1}
\usepackage{xcolor, graphicx, ulem}
\usepackage{amsmath, amssymb}
\usepackage[colorlinks=true,urlcolor=blue,citecolor=blue,linkcolor=blue]{hyperref}

\begin{document}

\title{Time-dependent quantum correlations in phase space}

\author{F. Krumm}\email{fabian.krumm@uni-rostock.de}\affiliation{Arbeitsgruppe Theoretische Quantenoptik, Institut f\"ur Physik, Universit\"at Rostock, D-18059 Rostock, Germany}
\author{W. Vogel}\affiliation{Arbeitsgruppe Theoretische Quantenoptik, Institut f\"ur Physik, Universit\"at Rostock, D-18059 Rostock, Germany}
\author{J. Sperling}\affiliation{Clarendon Laboratory, University of Oxford, Parks Road, Oxford OX1 3PU, United Kingdom}

\begin{abstract} 
	General quasiprobabilities are introduced to visualize time-dependent quantum correlations of light in phase space. 
	They are based on the generalization of the Glauber-Sudarshan $P$ function to a time-dependent $P$ functional [W. Vogel, Phys. Rev. Lett. {\bf 100}, 013605 (2008)],
	which fully describes temporal correlations of radiation fields on the basis of continuous phase-space distributions.
	This approach is nontrivial, as the $P$ functional itself is highly singular for many quantum states and nonlinear processes. 
	In general, it yields neither a well-behaved nor an experimentally accessible description of quantum stochastic processes.
	Our regularized version of this multitime-dependent quasiprobability is a smooth function and
	applies to stronger divergences compared to the single-time and multimode scenario.
	The technique is used to characterize an optical parametric process with frequency mismatch and a strongly nonlinear evolution of the quantized center-of-mass motion of a trapped ion.
	A measurement scheme, together with a sampling approach, is provided which yields direct experimental access to the regularized $P$ functional from measured data.
\end{abstract}


\date{\today}
\maketitle

\section{Introduction}\label{Sec:Introduction}

	Nonclassical effects, such as photon antibunching \cite{KDM77}, squeezing \cite{W83,Slusher,Wu,Va08,Va16}, and entanglement \cite{EPR35,S35}, have been known for many decades. Their verification, classification, quantification,
	and application remain challenging tasks of modern quantum optics.
	For the distinction of genuine quantum interferences from classical optical effects, two major techniques have been established.
	The first one is the application of various types of nonclassicality criteria based on observable quantities.
	The second one is the investigation of different kinds of phase-space distributions.
	The implementation of each of these techniques brings along its own characteristic advantages and challenges.

	A variety of nonclassicality criteria are suitable for different applications, depending on the quantum system or effect under consideration.
	Some of them consist of an infinite hierarchy of necessary and sufficient nonclassicality probes.
	Examples are criteria based on characteristic functions, moments, and their combination \cite{RiVo02,ShVo05,RSAMKHV15}.
	Yet, the full characterization of quantum effects requires the study of all orders of these hierarchies. 
	Even though this is impossible in general, these methods provide a plethora of sufficient nonclassicality conditions to successfully identify various types of quantum effects; cf. Ref. \cite{Miranowic10} for an overview.
	However, low-order criteria may fail to uncover the nonclassical character of particular quantum states.

	In such cases, the investigation of phase-space distributions may be advantageous. 
	Prominent examples are the Husimi $Q$ function \cite{H40}, the Wigner function \cite{W32}, the Glauber-Sudarshan $P$ function \cite{S63,G63}, their unification in terms of $s$-parameterized quasiprobability distributions \cite{CG69}, and the general distributions introduced by Agarwal and Wolf \cite{AgWo}.
	Nonclassicality is commonly defined via comparison of such quasiprobabilities with their classical counterparts. 
	The widely accepted definition of nonclassicality by Titulaer and Glauber relies on the $P$ function \cite{TG65,M86}:
	Whenever $P$ cannot be interpreted as a classical probability density, i.e., when it contains negativities, the state is referred to as nonclassical.
	Hence, this very quasiprobability will be our benchmark to identify quantum effects.
	However, the study of the $P$ function can become a cumbersome task, as it is highly singular for many quantum states \cite{S16}.
	In general, an experimental reconstruction of the $P$ function is only possible if a proper regularization procedure is introduced \cite{KV10}. 
	Based on such a technique, one can also implement a direct sampling approach for the regularized $P$ functions \cite{ASVKMH15}, which yields direct access to the full information on general quantum states.

	The $P$ function itself represents the full information on the quantum state $\hat \rho$ of a radiation mode at an arbitrary but fixed time in a diagonal representation,
	\begin{equation}
		\label{Eq:SinglePfunction}
		\hat \rho  =\int d^2\alpha P(\alpha) |\alpha \rangle \langle \alpha |,
	\end{equation}	
	with coherent states $|\alpha \rangle$ with complex amplitudes $\alpha$.
	Yet, correlations between multiple points in time (i.e., temporal correlations) play a fundamental role in quantum optics and quantum information theory. 
	For instance, the early demonstration of the quantum nature of light via photon antibunching was based on the detection of two-time intensity correlation properties \cite{KDM77}.
	Also, the photocounting theory depends on multitime correlation functions of the light field to be measured \cite{KK64}.
	It is noteworthy that the corresponding field correlations are normal and time ordered.
	In particular, this is relevant when the interaction dynamics is described by an explicitly time-dependent Hamiltonian or when the radiation field is emitted by atomic sources. 
	Recently, the Keldysh-ordered full counting statistics has been studied in the context of negativities of quasiprobabilities \cite{HC16,H17}. 

	Early studies of parametric processes in terms of time-dependent correlation functions were reported by Mollow \cite{Mollow73}.
	Temporal correlations have also been studied in the quantum dynamical theory of the fluctuations in a degenerate optical parametric oscillator \cite{PW94}.
	More recently, the multimode parametric dynamics has been further investigated by including time-ordering effects \cite{CBMS13}.
	For an explicitly time-dependent parametric interaction, temporal quantum correlations have also been considered in terms of characteristic functions \cite{KSV16}.

	Another field where time ordering is important is the passive filtering of light emitted from atomic sources.
	A general theory for the effects of passive optical systems on quantum light was developed in Ref. \cite{KVW87}.
	In this context, a careful treatment of non-equal-time commutators is crucial.
	Although such commutation rules are not explicitly known in general, their effects can be handled in a closed form; for details see Chap. 2.7 in Ref. \cite{VW06}.
	Yet another example stems from the spectral filtering of quantum light from atomic sources for which spectral squeezing \cite{KVW86} and spectral intensity correlations \cite{Knoell86,Cresser87} of the atomic resonance fluorescence have been studied to some extent.

        Temporal correlations are also considered in the framework of the Leggett-Garg inequalities \cite{LG85,ELN14}---sometimes referred to as \textit{temporal Bell inequalities}.
	Recently, the latter were extended to continuous-variable systems placed in a squeezed state \cite{MV16}.
	It turns out that the application areas of temporal correlation properties of radiation fields are a wide-ranging field of research \cite{KK64,Mollow73,PW94,CBMS13,KSV16,KVW87,VW06,KVW86,Knoell86,Cresser87,LG85,ELN14,MV16,BKetal15,arxNMR,cpetal15}.
	Hence, a complete characterization of such correlations is a subject of broad interest.
	
	Also, for applications of time-dependent quantum correlations of light in quantum technology, a full characterization of such complex quantum effects is indispensable.
	For this purpose, a space-time-dependent phase-space representation has been introduced by generalizing the Glauber-Sudarshan $P$ function to a space-time-dependent $P$ functional \cite{V08}.
	This functional renders it possible to formulate an infinite set of nonclassicality conditions in terms of normal-
	and time-ordered field correlation functions, which are accessible by homodyne correlation measurements \cite{SV06}.
	However, such a verification of quantum correlations is hardly used, as it requires the detection of a manifold of correlation functions.

	In conclusion, a direct study of the $P$ functional would be favorable and would lead to a deep and general understanding of temporal quantum correlations. 
	However, even for a single time, the $P$ functional can be highly singular.
	Even more severely, the singularities of the multitime $P$ functional are even not clearly understood yet.
	Although it is well known that the singularities of the Glauber-Sudarshan $P$ function are caused by the normal-ordering prescription, it is---to our best knowledge---presently unknown whether or not the time-ordering prescription, 
	  occurring in the $P$ functional, can give rise to singularities stronger than those of a multimode (but single-time) $P$ function.
	In our recent contribution \cite{KSV16}, we formulated nonclassicality tests to uncover time-dependent quantum effects.
	This method was based on the characteristic function, i.e., the Fourier transform of the $P$ functional, which is indeed a regular function.
	Still, until now, a proper regularization procedure for the $P$ functional itself has not been established for the multitime case.
	Such a method, however, would be important for a comprehensive understanding and potential applications of general quantum correlations of radiation fields.
	Here, it is also noteworthy that the general quantum correlations under study even include entanglement as a subset \cite{SV09}. 

	In the present paper, we develop a rigorous formalism that describes nonclassical multitime correlations in terms of smooth nonclassicality quasiprobabilities.
	The regularity of our phase-space function applies to any evolution of the optical system.
	This enables us to verify quantum effects through the negativity of our quasiprobability representation for quantum states of general radiation fields.
	For example, we apply our method to an explicitly time-dependent parametric process and to a nonlinear dynamics of the quantized motion of a trapped ion.
        It is shown that the multitime scenario implies singularities of the $P$ functional stronger than those of the single-time multimode $P$ function. Even those singularities are suppressed by our regularization technique. 
	Eventually, we formulate the measurement theory for the direct sampling of the regularized $P$ functional, which allows for an experimental visualization of multitime quantum correlations of light in phase space.

	The paper is structured as follows. 
	Section \ref{Sec:NC} recapitulates the concept of single- and multitime nonclassicality. 
	In Sec. \ref{Sec:QuasiProb}, we formulate the regularization procedure of the multitime-dependent $P$ functional.
        Afterwards, in Sec. \ref{Sec:Application} we apply the introduced techniques to an optical parametric process with frequency mismatch. 
        The nonlinear evolution of a laser-driven trapped ion is analyzed in Sec \ref{Sec:TrappedIon}, which includes more complex time-dependent commutation rules and strongly enhanced singularities of the $P$~functional.
	In Sec. \ref{Sec:Measurement}, we propose an experimental scheme for efficiently measuring the quantities under study.
	Finally, a summary and some conclusions follow in Sec. \ref{Sec:Conclusions}.

\section{Nonclassicality}\label{Sec:NC}

\subsection{Single-time nonclassicality filters}

	In the single-time scenario, the $P$ function can be used to express any quantum state as a formal mixture of coherent states \cite{G63,S63,TG65,M86,VW06}; cf. Eq.~(\ref{Eq:SinglePfunction}).
	However, it is due to the singular behavior of $P$, which occurs for many quantum states, such as Fock or squeezed states, that a direct experimental access to this distribution is impossible.
	The singularity is equivalent to an unbounded characteristic function $\Phi$, being the Fourier transform of $P$.

	To resolve the issue of singularities and to identify the nonclassicality of quantum states in experiments through negativities of properly defined quasiprobabilities, a regularization procedure was introduced \cite{KV10}.
	The resulting regularized $P$ function is consequently defined via the Fourier transform $\mathcal{F}$ or convolution $\ast$,
	\begin{equation}\label{Eq:SingTimeFilt}
		P_\Omega(\alpha)=\mathcal{F}_\beta[\Omega_w(\beta)\Phi(\beta)](\alpha)=(P \ast \tilde{\Omega}_w)(\alpha) ,
	\end{equation}
	with the filter function $\Omega_w$ or its inverse Fourier transform $\tilde{\Omega}_w$ and both depending on a width parameter $w>0$.
	This, in general, non-Gaussian filter function needs to satisfy several conditions \cite{KV10}:
	\begin{enumerate}
		\item $\Omega_w(\beta) \Phi(\beta)$ is rapidly decaying for all (finite) values $w$.
		This is necessary to assure that the regularized $P$ function exists and is even smooth for all states and for all filter widths (cf. also Ref. \cite{ASV13}, Appendix A).

		\item The Fourier transform $\tilde \Omega_w$ is a probability density.
		Especially, its nonnegativity is important, as we want to visualize the negativities of the original $P$ function,
		which define the nonclassicality.
		Thus, the filter must not contribute negativities.

		\item The limit $\lim_{w \rightarrow \infty} \Omega_w(\beta) =1$ assures that the original $P$ function is recovered for $w\to\infty$.
	\end{enumerate}

	The regularized $N$-mode and single-time $P$ function is obtained from the generalization \cite{ASV13}
	\begin{equation}
		\label{Eq:multi-modeConvolution}
		P_\Omega(\boldsymbol \alpha)=\mathcal{F}_{\boldsymbol \beta}[\Omega_w(\boldsymbol\beta)\Phi(\boldsymbol \beta)](\boldsymbol\alpha),
	\end{equation}
	where $\boldsymbol \alpha,\boldsymbol \beta\in\mathbb C^N$.
	One possibility for constructing a multimode filter is a product of single-mode filters,
	$\tilde \Omega_w (\boldsymbol \alpha)=\prod_{j=1}^N\tilde \Omega_w (\alpha_j)$.
	Note that the multimode characteristic function $\Phi$ is, in general, unbounded, $\sup_{\boldsymbol \beta\in\mathbb C^N} |\Phi(\boldsymbol \beta)|=\infty$, but can be bounded through a diverging function,
	\begin{equation}
		\label{Eq:multi-mode}
		|\Phi(\boldsymbol \beta)  |  \leq  \exp [ |\boldsymbol{\beta} |^2/2],
	\end{equation}
	which can be easily derived:
	Using the definition of the multimode characteristic function and the displacement operators $\hat D(\beta_j)=\exp[\beta_j \hat a_j^\dag -\beta_j^\ast \hat a_j]$ with $\boldsymbol \beta=(\beta_1, \dots, \beta_N)^{\rm T}$, one obtains
	\begin{equation}
		\label{Eq:multi-mode2}
		|\Phi(\boldsymbol\beta)| {=} \left| \left\langle {:} \prod_{j=1}^N \hat D( \beta_j) {:} \right\rangle \right|
		 {=} \left|\left\langle e^{ \sum_{j=1}^N \beta_j \hat{a}^\dag_j  } e^{- \sum_{j=1}^N \beta_j^\ast \hat a_j  }\right\rangle\right|,
	\end{equation}
	where $\hat a_l$ labels the annihilation operator of the $l$th radiation mode and $: \dots :$ connotes the normal ordering prescription.
	That is, all creation operators are placed to the left of the annihilation operators without making use of the bosonic commutation relations.
	As $[\hat a_i,\hat a_j]=0$ and $[\hat a_i,\hat a_j^\dag]=\delta_{ij}$ hold ($\delta$ is the Kronecker symbol), one can use the standard (i.e., first-order) Baker-Campbell-Hausdorff (BCH) formula to obtain
	\begin{equation}
		|\Phi(\boldsymbol\beta)|= \left| \left\langle \prod_{j=1}^N \hat D( \beta_j) \right\rangle\right|e^{ |\boldsymbol{\beta} |^2/2}.
	\end{equation}
	As for the unitary displacement operators $\hat D(\beta_j)$ it holds that $\| \prod_{j=1}^N \hat D( \beta_j)\|\leq 1$, and one readily verifies Eq. \eqref{Eq:multi-mode}.
	This estimation also holds true for any time evolution of a quantum state $\hat \rho(t)$ for a single time $t$.
	Hence, one finds that the slope of the characteristic function of the $N$-mode $P$ function is bounded by an inverse Gaussian factor, which also bounds the singularities of the multimode $P$ function \cite{S16}.
	However, the regularization procedure so far recapitulated is restricted to single-time properties of a quantum system.

\subsection{Multitime $P$ functional}

	In the more general multitime scenario, the situation is very different.
	It is nontrivial to give a similar expansion as in Eq. \eqref{Eq:SinglePfunction}, because it is a cumbersome task to define the corresponding multitime density matrix \cite{APT14}.
	Thus, one needs a generalized, multitime-dependent version of the $P$ function \cite{V08}.
	We will discuss this concept in the continuation of this section.
	The resulting $P$ functional is formulated by using normal- and time-ordered expressions which are accessible in quantum correlation measurements \cite{VW06,SV06}.
	They also occur in the photocounting theory \cite{KK64} whenever source fields play a significant role in the description of a quantum state of light.

	Let us consider an observable $\hat O[ \{\hat a(t_i)\}_{i=1}^k]$, which depends on the bosonic annihilation operators $\hat a(t_i)$ and 
	creation operators $\hat a(t_i)^\dag$ (not explicitly written as an argument in $\hat O$) at arbitrarily chosen points in time, $t_1 \leq \dots \leq t_k$.
	Throughout this work, $k$ denotes the number of different points in time.
	For simplicity, we treat the case of a single spatial-frequency optical mode but $k$ nonmonochromatic (temporal) modes.
	The extension to $N$ spatial-frequency modes is straightforward, via $\hat a^{(\dag)} \rightarrow (\hat a^{(\dag)}_1,\dots,\hat a^{(\dag)}_N)^{\rm T}$, as exemplified in the previous subsection for a single time.
	The definition of the singular $P$ functional was introduced in terms of space-time-dependent field operators in Ref. \cite{V08}, which already includes the most general scenario.
	It is a function of the $k$ coherent amplitudes at the considered $k$ points in time, $P[\alpha_1,\dots,\alpha_k;t_1,\dots ,t_k]$.

	Using this $P$ functional, a multitime-dependent expectation value of the given observable $\hat O$ may be written as
	\begin{eqnarray}
		\nonumber
		\big\langle \begin{smallmatrix}\circ \\ \circ\end{smallmatrix} \hat O[ \{\hat a(t_i)\}_{i=1}^k] \begin{smallmatrix}\circ \\ \circ\end{smallmatrix} \big\rangle
		=& &
		\int {d}^2\alpha_1 \dots \int {d}^2\alpha_k O(\alpha_1,\dots,\alpha_k)
		\\& &\times  P[\alpha_1,\dots,\alpha_k;t_1,\dots ,t_k],
		\label{Eq:OrderesExpVal}
	\end{eqnarray}
	where we omitted the dependence on the complex conjugated variables and operators.
	The symbol $\begin{smallmatrix}\circ \\ \circ\end{smallmatrix}  \dots \begin{smallmatrix}\circ \\ \circ\end{smallmatrix} = \mathcal{T} :\dots :$ represents the normal ($: \dots :$)- and time ($\mathcal T$)-ordering prescription.
	Namely, the operators in Eq. \eqref{Eq:OrderesExpVal} have to be normal ordered---creation operators to the left of annihilation operators---and then time ordering is performed, i.e., the time-dependent creation (annihilation) operators are sorted 
	with increasing (decreasing) time arguments from left to right \cite{VW06}.
	Note that from the theory of photoelectric detection of light it is well
	known that observable correlation functions are subjected to normal and time ordering.
	An example of such a function is the second-order intensity correlation function, $g^{(2)}$, which corresponds to the expectation value of the operator ${\hat O \sim \hat a^\dag(t)  \hat a^\dag(t + \Delta t) \hat a(t + \Delta t)\hat a(t) }$.
	This quantity is essential for the verification of photon antibunching \cite{KDM77,P82}.

	From the general structure of Eq. \eqref{Eq:OrderesExpVal}, we can observe that the $P$ functional has formally the meaning of a joint probability distribution of the coherent amplitudes $\alpha_i\equiv \alpha(t_i)$ at $k$ points in time.
	We use the term ``formally'' here, as $P$, in general, does not fulfill all the properties of a probability density in the sense of classical stochastics.
	This means that the $P$ functional is a joint quasiprobability, defined as the quantum expectation value \cite{V08}
	\begin{equation}
		\label{Eq:Pfunctional}
		P[\{\alpha_i;t_i\}_{i=1}^k]=\Big\langle  \begin{smallmatrix}\circ \\ \circ\end{smallmatrix} \prod_{i=1}^k \hat \delta(\hat a(t_i)-\alpha_i)  \begin{smallmatrix}\circ \\ \circ\end{smallmatrix} \Big\rangle,
	\end{equation}
	where $\hat \delta$ denotes the operator-valued $\delta$ distribution.
	To clarify the terms, let us stress the following: The notion $P$ function is used for characterizing the quantum state at a single (arbitrary but fixed) time.
	The notion $P$ functional, on the other hand, applies when field amplitudes including their time dependencies are relevant.
	This also means that in the case $k=1$, the time-ordering prescription becomes meaningless, $ \begin{smallmatrix}\circ \\ \circ\end{smallmatrix} \cdots  \begin{smallmatrix}\circ \\ \circ\end{smallmatrix} \rightarrow {:}\,\cdots\, {:} $.
	In this scenario, the Glauber-Sudarshan $P$ function in Eq. \eqref{Eq:SinglePfunction} is recovered.

	The classicality (nonnegativity) of the multitime functional, \eqref{Eq:Pfunctional}, leads to a hierarchy of classical inequalities in term of moments.
	Their violation certifies general quantum correlations of light \cite{V08}. 
	Special cases were also studied in Ref. \cite{O07} for characterizing two-photon quantum interferences.

\section{The filtered $P$ functional}\label{Sec:QuasiProb}

	Starting from the definition \eqref{Eq:Pfunctional}, one can always express the $P$ functional through its characteristic function $\Phi$,
	\begin{equation}
		\label{Eq:FourierFunctional}
		P[\{\alpha_i;t_i \}_{i=1}^k] =\mathcal{F}_{\{ \beta_i\}_{i=1}^k}  [ \Phi(\{\beta_i ;t_i \}_{i=1}^k) ](\{\alpha_i\}_{i=1}^k),
	\end{equation}
	where $\mathcal{F}_{\{ \beta_i\}_{i=1}^k} =\mathcal{F}_{\beta_1} \cdots \mathcal{F}_{\beta_k}$ is a product of Fourier transforms for the $k$ different degrees of freedom.
	Let us recall that the (time-dependent) operator-valued $\delta$ distribution is defined as the Fourier transform of the (time-dependent) displacement operator,
	${\hat \delta(\hat a(t)-\alpha)=\mathcal{F}_\beta[\hat D(\beta;t)](\alpha)}$, with $\hat D(\beta;t)=\exp \left[ \beta \hat a(t)^\dag - \beta^\ast \hat a(t) \right] $.
	Hence, one readily gets that
	\begin{eqnarray}
		\label{Eq:multi-timeCF}
		& &\Phi(\{\beta_i ;t_i \}_{i=1}^k)=\left\langle  \begin{smallmatrix}\circ \\ \circ\end{smallmatrix} \prod_{i=1}^k \hat D(\beta_i;t_i)  \begin{smallmatrix}\circ \\ \circ\end{smallmatrix} \right\rangle  \\
		&=& \left\langle   \prod_{i=1}^k e^{\beta_i \hat a^\dag(t_i)}  \prod_{i=1}^k e^{-\beta_{k+1-i}^\ast \hat a(t_{k+1-i})}  \right\rangle, \nonumber
	\end{eqnarray}
	which is the multitime-dependent characteristic function (MTCF) of the $P$ functional for ${t_1 \leq \dots \leq t_k}$ \cite{KSV16}.
	The operator product for arbitrary $\hat A(t)$ is defined as $\prod_{i=1}^k \hat A(t_i) = \hat A(t_1)  \dots  \hat A(t_k)$.
	Since the operators are in general not commuting, the operator products contain the time ordering from the first line in Eq. \eqref{Eq:multi-timeCF}.

	At first sight, the above expression, \eqref{Eq:multi-timeCF}, resembles that for the multimode characteristic function [Eq. \eqref{Eq:multi-mode2}]. 
	However, there are two significant differences when considering the multitime scenario:
	
	(i) One needs to consider non-equal-time commutators of the field operators, which are, in general, nonvanishing or not even proportional to unity \cite{VW06,C83,C84,KVW87}.
	This means that $[\hat a(t),\hat a(t')^\dag] \not \propto \hat 1$ for $t\neq t'$, which is a crucial point, as such commutators do not necessarily commute with other operators.
	Hence, the standard BCH formula is insufficient and higher-order terms need to be taken into account.
	Their calculation and the related convergence considerations are complex problems \cite{BC04}, which complicates the issue of finding a bound of the absolute square of the MTCF.
	This result is a major difference compared to the multimode (single-time) case [Eq. \eqref{Eq:multi-mode}].

	(ii) Resulting from the structure of the functional, \eqref{Eq:Pfunctional}, the time-ordering prescription (beside the normal-ordering prescription) has to be considered.
	As we saw in the derivation of Eq. \eqref{Eq:multi-mode}, the factor $e^{|\beta|^2/2}$ for a single radiation mode is caused by the normal ordering.
	The question arises whether or not the time ordering itself does cause a stronger rising behavior of the MTCF.
	
	Altogether, the MTCF may be a more strongly growing function of $\beta_1,\dots,\beta_k$ compared with the single-time but multimode scenario.
	However, this asymptotic behavior is important as it could lead to stronger singularities of the corresponding $P$ functional compared with the multimode $P$ function.
	This problem, to our best knowledge, has not been studied yet.
	Our rigorous regularization procedure in the multitime scenario has to include this eventuality.
	This also means that our approach, to be formulated, needs to be applicable to any dynamics of a quantum optical system. 
	
\subsection{Universal multitime regularization}\label{Subsec:UMR}

	As emphasized above, the singularities of $P$ in Eq. \eqref{Eq:FourierFunctional} are caused by the fact that the characteristic function $ \Phi(\{\beta_i ;t_i \}_{i=1}^k)$ [Eq. \eqref{Eq:multi-timeCF}] is, in general, unbounded.
	Hence, the integrals of the Fourier transforms in Eq. \eqref{Eq:FourierFunctional} do not converge.
	When generalizing the approach in Eq. \eqref{Eq:SingTimeFilt}, our multitime filter $\Omega_{\boldsymbol{w}}$ needs to assure the fast decay of the filtered characteristic function,
	\begin{equation}
		\Phi_{\Omega}(\{\beta_i ;t_i \}_{i=1}^k)
		= \Phi(\{\beta_i ;t_i \}_{i=1}^k) \Omega_{\boldsymbol{w}}(\{\beta_i ;t_i \}_{i=1}^k).
	\end{equation}
	Consequently, the regularized $P$ functional in terms of the Fourier transform is defined as
	\begin{equation} 
		\label{Eq:multi-timefiltering2}
		P_\Omega[\{\alpha_i;t_i \}_{i=1}^k] = \mathcal{F}_{\{ \beta_i\}_{i=1}^k}  [ \Phi_\Omega(\{\beta_i ;t_i \}_{i=1}^k)  ] (\{\alpha_i\}_{i=1}^k),
	\end{equation}
	where $\boldsymbol w$ denotes a tuple of width parameters that is specified later.

	In the following, let us formulate some simple observations.
	We consider a continuous function $\varphi(z)$ depending on a real-valued parameter $z$ that might diverge for $|z|\to\infty$.
	In addition, we employ the triangular function,
	\begin{equation}
		\mathrm{tri}(z)=\left\lbrace\begin{array}{ll}
			(1+z) & \text{ for } z\in[-1,0], \\
			(1-z) & \text{ for } z\in[0,1], \\
			0 &  \text{ else. }
		\end{array}\right.
	\end{equation}
	It holds that the product $\varphi(z)\mathrm{tri}(z)$ is bounded and continuous since the product of two continuous functions is continuous, and since the triangular function has the compact support $[-1,1]$, it is identical to $0$ for $|z|>1$.
	Further on, it is easy to check that the one-dimensional Fourier transform of $\mathrm{tri}(z)$ is a probability density.
	Rescaling the argument with $w>0$, $\mathrm{tri}(z/w)$, yields a rescaled probability density, with the support of $\mathrm{tri}(z/w)$ changing to the interval $[-w,w]$.
	In particular, for $w\to\infty$, this rescaled triangular function converges pointwise to the constant function $1$, for which the Fourier transform is a $\delta$ distribution.

	Returning to our initial filtering problem and keeping those observations in mind, we define the filter function
	\begin{equation} 
		\label{Eq:productfilter}
		\Omega_{\boldsymbol w} (\{\beta_i \}_{i=1}^k)=\prod_{i=1}^k\left(
			{\rm tri}({\rm Re}[\beta_i]/w_i) {\rm tri}({\rm Im}[\beta_i]/w_i)
		\right),
	\end{equation}
	using different filter parameters for each time, $\boldsymbol w = (w_1,\dots ,w_k)$, and ${\rm Re}[\beta]$ with ${\rm Im}[\beta]$ denoting the real and imaginary part of $\beta$, respectively.
	From our observation above it directly follows that requirements 1--3 (cf. Sec. \ref{Sec:NC}) for a filter are satisfied.
	In particular and due to its compact support, the filter, \eqref{Eq:productfilter}, suppresses any rising behavior of the MTCF.
	This also means that the filtered $P$ functional, \eqref{Eq:multi-timefiltering2}, exists always as a smooth function \cite{comment}.

\begin{figure}[tb]
	\centering
	{\includegraphics*[width=8.0cm]{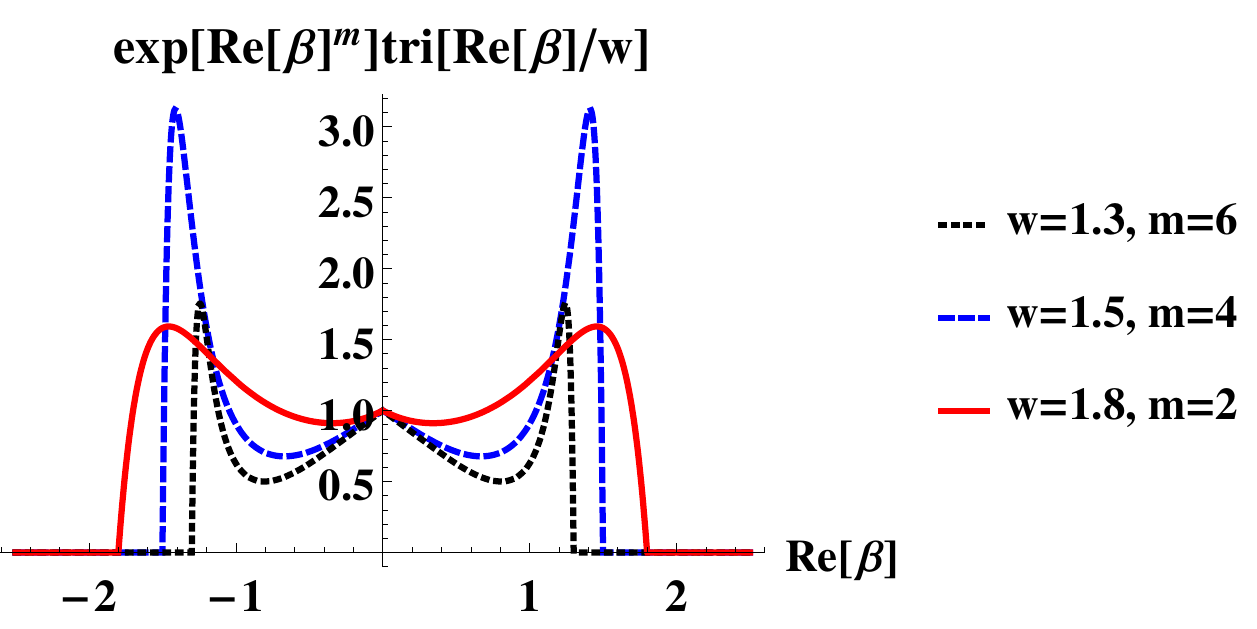}}
	\caption{
		Plot of the triangular function $\mathrm{tri}({\rm Re}[\beta]/w)$ multiplied by an exponential rising function ${\exp[|\beta|^m]}$.
		We see that the resulting function is bounded, and thus, its Fourier transform is a continuous function.
	}\label{Fig:trifilter}
\end{figure}

	For illustration, a plot of the filter, multiplied by a factor $\exp[{\rm Re}[\beta]^m]$ for $m=2,4,6$, is given in Fig. \ref{Fig:trifilter}.
	In the case $m=2$, the factor resembles the growth factor of a single-time-characteristic function \cite{P86}.
	The parameters $m=4,6$ correspond to faster increments that might result from non-equal-time commutation relations of nonlinear interactions.
	Note that a filter of this type is needed for multitime-dependent phenomena, as the behavior of the MTCF cannot be estimated in general. 
	The multitime commutation rules of the bosonic operators are unknown for general interaction problems; for details see Chap. 2.7 of \cite{VW06}.
	With our filters in Eq. \eqref{Eq:productfilter}, however, we ensure that the multitime nonclassicality quasiprobabilities are smooth functions and they can be directly sampled in experiments, which is shown in Sec. \ref{Sec:Measurement}.

\subsection{Discussion}

	Let us formulate some preliminary conclusions.
	The problem of regularizing the $P$ functional has been treated.
	Although the singularities of this multitime quasiprobability are unknown, we formulated a regularization approach via a compact filter function, \eqref{Eq:productfilter}, which applies to any nonlinear interaction dynamics of radiation fields.
	On this basis, the well-behaved quasiprobability, \eqref{Eq:multi-timefiltering2}, exists and it is negative for a width $w$ if and only if the light field under study is nonclassical---including multitime quantum correlations.
	Our applied filter suppresses any rising behavior of the MTCF for $|\beta_l|\to\infty$.
	This also includes scenarios where the slope of the MTCF increases more rapidly than an inverse Gaussian, which might result from the dynamics of complex interactions.

	Remarkably, a triangular filter has already been used for the single-mode and -time scenario \cite{KV10}.
	There, the compact support was considered a deficiency because parts of the characteristic function are multiplied by $0$ and, thus, do not contribute to the filtered $P$ function.
	Here, however, this compactness serves as a beneficial resource which renders it possible to filter a multitime $P$ functional.
	Moreover, similarly to the multimode scenario \cite{ASV13}, we have a filter in a product form, \eqref{Eq:productfilter}.
	It is worth mentioning that we could also and equivalently employ any filter with a compact support, including radial symmetric filters and higher-order autocorrelation-function filters \cite{KV12,KV14}.

	The basic definition of the multitime nonclassicality is the inability to interpret the singular $P$ functional, \eqref{Eq:Pfunctional}, as a joint probability distribution of a classical stochastic process \cite{V08}.
	Also in Ref. \cite{V08}, a hierarchy of quantum correlation conditions has been formulated on the basis of measurable space-time-dependent field correlation functions.
	An approach to formulating multitime nonclassicality tests on the basis of the MTCF was subsequently formulated as well \cite{KSV16}.
	Here, in contrast, we introduce regular nonclassicality quasiprobabilities that enable us to study quantum correlations between multiple points in time directly via the corresponding negativities of smooth phase-space distributions of quantum stochastic, optical processes.

\section{Parametric Processes}\label{Sec:Application}

	Let us now apply our approach to the characterization of temporal quantum effects on a specific physical system.
	In our recent work \cite{KSV16}, we studied the parametric process based on the MTCF.
	This and related parametric interactions are a fundamental tool for generating nonclassical light in modern experimental quantum optics, e.g., squeezed light \cite{Slusher,Wu,Va08,Va16}.
	Note that various single-photon sources are based on parametric down-conversion \cite{HO87,CS12,CD06,PJ05,RS04,LH01}.
	Due to the resulting wide range of applications of this process, let us reconsider this system from the perspective of the technique derived here.
	
	The effective Hamiltonian of the quantum system in the interaction picture is
	\begin{equation}
		\label{Eq:Hamiltonian}
		\hat H_{\rm int}(t)= \hbar \kappa ( \hat a^{\dag 2} e^{-i \delta t}+\hat a^2 e^{i \delta t} )
	\end{equation}
	with a positive frequency mismatch $\delta= \omega_{p}-2 \omega_{a}$ and with $\omega_{p}$ and  $\omega_{a}$ being the pump and signal frequency, respectively.
	In this process, a strong (classically described) pump field creates pairs of (equal-frequency) signal photons.
	Due to the violation of multitime-dependent classical inequalities, we have already demonstrated the presence of two-time quantum correlations \cite{KSV16} for this process.
	We also stress that the dynamical behavior exhibits a nontrivial dependence on time, as we have, in general, a nonvanishing commutator $[\hat H_{\rm int}(t),\hat H_{\rm int}(t')]\neq 0$ for different times $t$ and $t'$.
	In this section, let us focus on the visualization of quantum correlations directly in terms of negativities of the regularized $P$ functional, which has not been considered before.

	The coupled equations of motion of the signal field operators $\hat a$ and $\hat a^\dag$ read as
	\begin{eqnarray}\label{Eq:MatrixM} 
		\frac{d}{dt} \begin{pmatrix}
			\hat a(t) \\
			\hat a(t)^\dag
		\end{pmatrix}
		&=& \begin{pmatrix}
		0 & -2i \kappa e^{-i\delta t} \\
		2i \kappa e^{i\delta t}& 0
		\end{pmatrix}
		\begin{pmatrix}
			\hat a(t) \\
			\hat a(t)^\dag
		\end{pmatrix}.
	\end{eqnarray}
	After decoupling, one obtains second-order equations of motion,
	\begin{equation}
		\frac{d^2}{dt^2} \hat a(t)+ i \delta \frac{d}{dt}\hat a(t) - 4 \kappa^2 \hat a(t) =0.
	\end{equation}
	The solution can be found via standard algebra,
	\begin{equation}
		\label{Eq:ExactTO}
		\hat a(\tau)= u_{1}(\tau) \hat a + u_{2}(\tau) \hat a^\dag,
	\end{equation}
	where we have defined the following dimensionless quantities: $\vartheta_r=\pi \sqrt{16-r^2}/4$ (representing the eigenfrequency), $r=\delta/\kappa$ (the coupling ratio), $\tau=2 \kappa t/\pi$ (a time in ``natural'' units of the system),
	and the two functions
	\begin{eqnarray}
		\label{Eq:ExactSolution}
		u_1(\tau)&=& e^{-i \pi r \tau/4} \left[\cosh{(\vartheta_r \tau)} + \frac{i \pi r}{4\vartheta_r} \sinh{(\vartheta_r \tau)}\right],  \nonumber \\
		u_2(\tau)&=& \frac{-  i \pi}{\vartheta_r} e^{-i \frac{\pi}{4} r \tau}  \sinh{(\vartheta_r \tau)}.
	\end{eqnarray}
	The initial condition is $\hat a(\tau=0)\equiv \hat a$.

\subsection{Single-time scenario}\label{SubSec:STRegul}

	To clarify the filtering procedure and to demonstrate the applicability of the filter to the single-time dynamics of the system, we first consider the single-time scenario.
	In this case, $k=1$, the $P$ functional in Eq. \eqref{Eq:FourierFunctional} is obtained from the inverse Fourier transform, \eqref{Eq:multi-timeCF}:
	\begin{equation}
		P[\alpha^\prime;\tau]=\frac{1}{\pi^2}\int {d}^2\beta e^{\beta^{\prime \ast} \alpha^\prime-\beta^\prime \alpha^{\prime \ast}} \langle \mathord{:} \hat D(\beta^\prime;\tau)\mathord{:} \rangle.
	\end{equation}
	Since we treat the time evolution in the interaction picture, the impact of the free-field Hamiltonian $\hat H_0=\hbar \omega_{a} \hat a^\dag \hat a$ is not directly included in the parameters $u_l$.
	However, it only acts as a classical rotation in phase space,
	$\hat D(\beta^\prime;\tau) \mapsto \hat D(\beta^\prime e^{i\omega_{a} t };\tau)$, which can be ignored, as we can perform a transformation $\beta^\prime e^{i\omega_{a} t } = \beta$ (likewise, $ e^{i\omega_{a} t }  \alpha^\prime  = \alpha$ in the original
	phase space).
	Using Eq. \eqref{Eq:multi-timeCF} for $k=1$, inserting the solution of the time evolution, \eqref{Eq:ExactTO}, using the decomposition $\beta=\beta_r+i\beta_i$ ($\beta_r,\beta_i\in\mathbb R$), and reordering the terms using the BCH formula, we get
	\onecolumngrid
	\begin{eqnarray}
		\label{Eq:STCFnativecoeff}
		\langle \mathord{:} \hat D(\beta;\tau)\mathord{:} \rangle \equiv \Phi(\beta;\tau)
		= \exp{\left[ 
		\left(\begin{smallmatrix} \beta_r \\ \beta_i \end{smallmatrix}  \right)^{\rm T}
		\left( \begin{smallmatrix}  \frac{1}{2}(1-|u_2^\ast - u_1|^2 )&  {\rm Im}[u_1 u_2] \\  {\rm Im}[u_1 u_2] &   \frac{1}{2}(1-|u_2^\ast + u_1|^2 ) \end{smallmatrix} \right)
		\left(\begin{smallmatrix} \beta_r \\ \beta_i \end{smallmatrix} \right) \right]}. 
	\end{eqnarray}
	\twocolumngrid
	Here, we have supposed that the initial state is the vacuum state and omitted to write the explicit time dependence, $u_l=u_l(\tau)$.

	To simplify the integration, we diagonalize the coefficient matrix in Eq. \eqref{Eq:STCFnativecoeff}.
	The transformation matrix $\boldsymbol S$, containing the normalized eigenvectors $\boldsymbol \chi_l$ ($l=+,-$), reads
	\begin{equation}
		\boldsymbol S=(\boldsymbol \chi_+, \boldsymbol \chi_-)=  \begin{pmatrix} \frac{-b_+}{\sqrt{1+|b_+|^2}} & \frac{-b_-}{\sqrt{1+|b_-|^2}} \\ \frac{1}{\sqrt{1+|b_+|^2}} & \frac{1}{\sqrt{1+|b_-|^2}} \end{pmatrix},
	\end{equation}
	with $b_\pm=(-{\rm Re}[u_1 u_2] \pm |u_1 u_2|)/{\rm Im}[u_1 u_2] \in \mathbb{R}$.
	The obtained normal coordinates are described via the classical rotation $ (\beta_r , \beta_i)^{\rm T} = \boldsymbol S   (\gamma_r , \gamma_i)^{\rm T} $ and we find
	\begin{eqnarray}
		& & \Phi(\gamma;\tau)= \exp{\left[ 
		\left(\begin{smallmatrix} \gamma_r \\ \gamma_i \end{smallmatrix}  \right)^{\rm T}
		\left( \begin{smallmatrix}  -c_+& 0 \\  0 &   -c_-\end{smallmatrix} \right)
		\left(\begin{smallmatrix} \gamma_r \\ \gamma_i \end{smallmatrix} \right) \right]},
		  \\ & & \text{with } c_\pm=  -\frac{1}{2}(1-|u_2^\ast-u_1|^2) \pm 2\frac{|u_1 u_2 |}{1+b_\pm^2}. \nonumber 
	\end{eqnarray}
	Including the free-field propagation, we further obtain
	\begin{eqnarray}
		P[\alpha;\tau]=& & \frac{1}{\pi^2}\int_{-\infty}^\infty {d} \gamma_r \int_{-\infty}^\infty {d} \gamma_i
		\\& & \times\nonumber
		\exp{\left[ 2i  A_r \gamma_r - 2i   A_i\gamma_i-c_+\gamma_r^2-c_-\gamma_i^2 \right]},
	\end{eqnarray}
	with 
	\begin{eqnarray}
		A_r&=&-(1+b_+^2)^{-1/2}({\rm Re}[\alpha]+b_+{\rm Im}[\alpha]), \\
		A_i&=&(1+b_-^2)^{-1/2}({\rm Re}[\alpha]+b_-{\rm Im}[\alpha]). \nonumber 
	\end{eqnarray}
	The application of our regularization procedure, \eqref{Eq:multi-timefiltering2}, to a single point in time yields
	\begin{eqnarray}
		\label{Eq:Psingletimecalc2}
		P_\Omega[\alpha;\tau]=& &\int_{-\infty}^\infty \frac{{d} \gamma_r}{\pi}   e^{ 2i  A_r  \gamma_r -c_+\gamma_r^2 } {\rm tri} ( \gamma_r/w)  \\
		 & &\times \int_{-\infty}^\infty \frac{{d} \gamma_i}{\pi}  e^{2i  (-A_i) \gamma_i-c_-\gamma_i^2 } {\rm tri} ( \gamma_i/w).  \nonumber 
	\end{eqnarray}
	The integral can be simplified (cf. Appendix C in \cite{S16}) via defining the function
	\begin{eqnarray}
		\label{Eq:Tfunction}
		T(y,g)
		&=& {\rm Re}\left[ \frac{2}{\pi} \int_0^1 dz e^{-gz^2+2iyz}(1-z) \right],
	\end{eqnarray}
	which relates to complex error functions, and we finally arrive at
	\begin{equation}
		P_\Omega[\alpha;\tau]=w^2 T(wA_r,w^2c_+)T(-wA_i,w^2c_-).
	\end{equation}

\begin{figure}[tb]
	\centering
	{\includegraphics*[width=8.0cm]{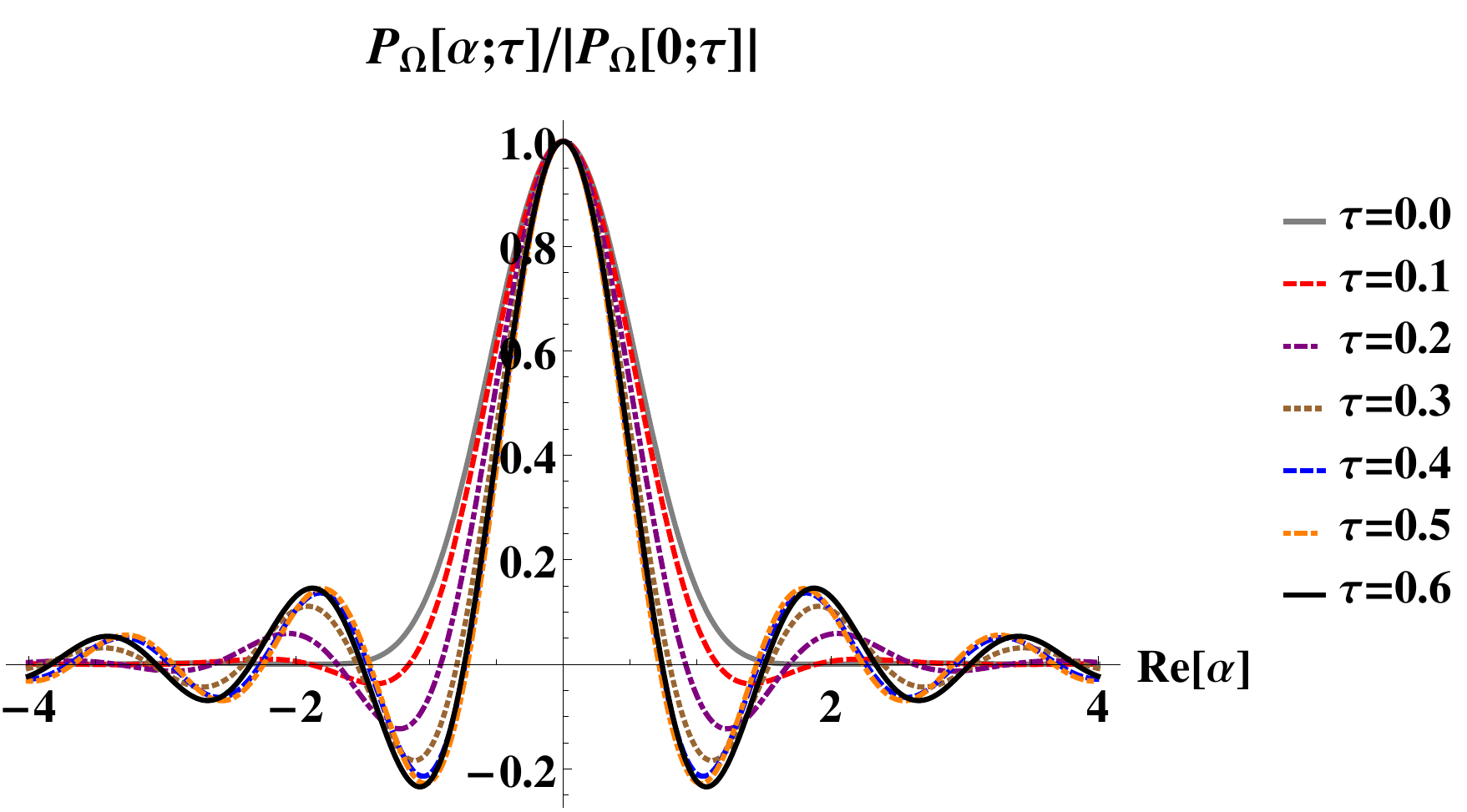}}
	\caption{
		Plot of the regularized and scaled phase-space distribution $ P_\Omega[\alpha;\tau]$ for several system times $\tau$.
		We chose ${\rm Im}[\alpha]=0$, $w=2.3$ for the triangular filter and $r =\delta/\kappa=10/\pi \approx 3.18$.
		The negativities for $\tau>0$ clearly show the evolution of the nonclassicality.
	}\label{Fig:PDCSingletimeP}
\end{figure}

	The temporal evolution in terms of $P_\Omega[\alpha;\tau]$ is shown in Fig. \ref{Fig:PDCSingletimeP} for different times $\tau=2\kappa t/\pi$.
	The negativities clearly display the nonclassicality of the system in terms of a regular and time-dependent quasiprobability.

\subsection{Singularities due to explicit time dependence}\label{SubSec:SingulTime}

	Let us now extend our studies to the more general multitime case.
	As discussed earlier, the main features of multitime correlations are due to 
	(i) non-equal-time commutation relations and
	(ii) the time-ordering prescription; see the beginning of Sec. \ref{Sec:QuasiProb}.
	The commutators for (i) can be straightforwardly computed for the system under study by using the exact solution, \eqref{Eq:ExactTO},
	\begin{eqnarray}
	\label{Eq:MTcommrel}
		& & [\hat a(\tau),\hat a(\tau+\Delta \tau)]\\
		&=& \det \begin{pmatrix}
				u_1(\tau) & u_2(\tau) \\
				u_1(\tau+\Delta \tau) &
				u_2(\tau+\Delta \tau)
		\end{pmatrix}\hat 1,  \nonumber 
	\end{eqnarray}
	with $u_l(\tau)$ given in Eq. \eqref{Eq:ExactSolution}.
	The effect of the time-ordering prescription in Eqs. \eqref{Eq:Pfunctional} and \eqref{Eq:multi-timeCF} can be analyzed in terms of the ratio
	\begin{equation}
		\label{Eq:ratio}
		\tilde \Phi = \left| \frac{\langle \mathord{\begin{smallmatrix}\circ \\ \circ\end{smallmatrix} } 
		\hat D(\beta_1,\beta_2;\tau,\tau+\Delta \tau)\mathord{\begin{smallmatrix}\circ \\ \circ\end{smallmatrix} } \rangle}{\langle \mathord{:} 
		\hat D(\beta_1,\beta_2;\tau,\tau+\Delta \tau)\mathord{: } \rangle} \right|,
	\end{equation}
	which relates the time- and normal-ordered quantities to solely normal-ordered ones.
	Since all appearing commutators are multiples of the identity, we get
	\begin{eqnarray}
		\tilde \Phi \ & & = |\exp \{ - \beta_1^\ast \beta_2^\ast[ \hat a (\tau), \hat a(\tau+\Delta \tau) ] \} | \\
		& & = \exp \{ - |\beta_1 \beta_2 | {\rm Re}[ e^{-i(\varphi_{\beta_1} + \varphi_{\beta_2}) }  [ \hat a (\tau), \hat a(\tau+\Delta \tau) ]  ] \} , \nonumber
	\end{eqnarray}
	where $\varphi_{\beta_j}=\arg\beta_j$ for $j=1,2$.
	Using the solutions, \eqref{Eq:ExactSolution}, one finds for $|r|\leq 4$ and $\vartheta_r=\pi \sqrt{16-r^2}/4$ that
	\begin{eqnarray}
		\label{Eq:Torisingfactor}
		& &{\rm Re}[ e^{-i(\varphi_{\beta_1} + \varphi_{\beta_2}) }  [ \hat a (\tau), \hat a(\tau+\Delta \tau) ]  ] \\
		&=&-\frac{\pi}{\vartheta_r} \sin\left[\frac{\pi r}{4} (2 \tau + \Delta \tau)+\varphi_{\beta_1}+\varphi_{\beta_2}\right] \sinh[\vartheta_r\Delta \tau]\hat 1. \nonumber
	\end{eqnarray}
	In other words, the influence of the time-ordering prescription is an additional term proportional to $\exp  \{\pm |\beta_1 \beta_2|\}$, where the sign depends on the 
	sine ($\sin$) term in Eq. \eqref{Eq:Torisingfactor}.
	As the hyperbolic sine ($\sinh$) is monotonically increasing, the strength of this factor increases with $\Delta \tau$.
	For the $P$ functional (i.e., performing a Fourier transformation), this factor increases (for $\exp  \{+ |\beta_1 \beta_2|\}$) or decreases (for $\exp  \{- |\beta_1 \beta_2|\}$) the strength of the singularities.

	Let us apply our filter procedure introduced in Sec. \ref{Subsec:UMR}.
	We use identical filter widths, $w_1=w_2=w$ [cf. Eqs. \eqref{Eq:multi-timefiltering2} and \eqref{Eq:productfilter}].
	The general procedure to compute $P_\Omega[\alpha_1,\alpha_2;\tau_1,\tau_1]$ is a straightforward extension of the one presented in Sec. \ref{SubSec:STRegul}.
	After some algebra, we get the two-time regularized $P$ functional,
	\begin{eqnarray}
		\label{Eq:Pfunctionmulti-timeRegul}
			 & & P_\Omega[\alpha_1,\alpha_2;\tau_1,\tau_1]
			\\ = & &  w^4 T(w \frac{f_{10}}{2i},-w^2 f_{20}) T(w \frac{f_{01}}{2i},-w^2 f_{02})
			\nonumber \\ & &\times T(w \frac{d_{10}}{2i},-w^2 d_{20})T(w \frac{d_{01}}{2i},-w^2 d_{02}), \nonumber 
	\end{eqnarray}
	with the definition of $T$ in Eq. \eqref{Eq:Tfunction}.
	The coefficients $f_{mn}$ and $d_{mn}$ together with a proper rotation of phase space can be obtained numerically as described in details in Sec. \ref{SubSec:STRegul} and generalized to four dimensions.

\begin{figure}[tb]
	\centering
	{\includegraphics*[width=8.0cm]{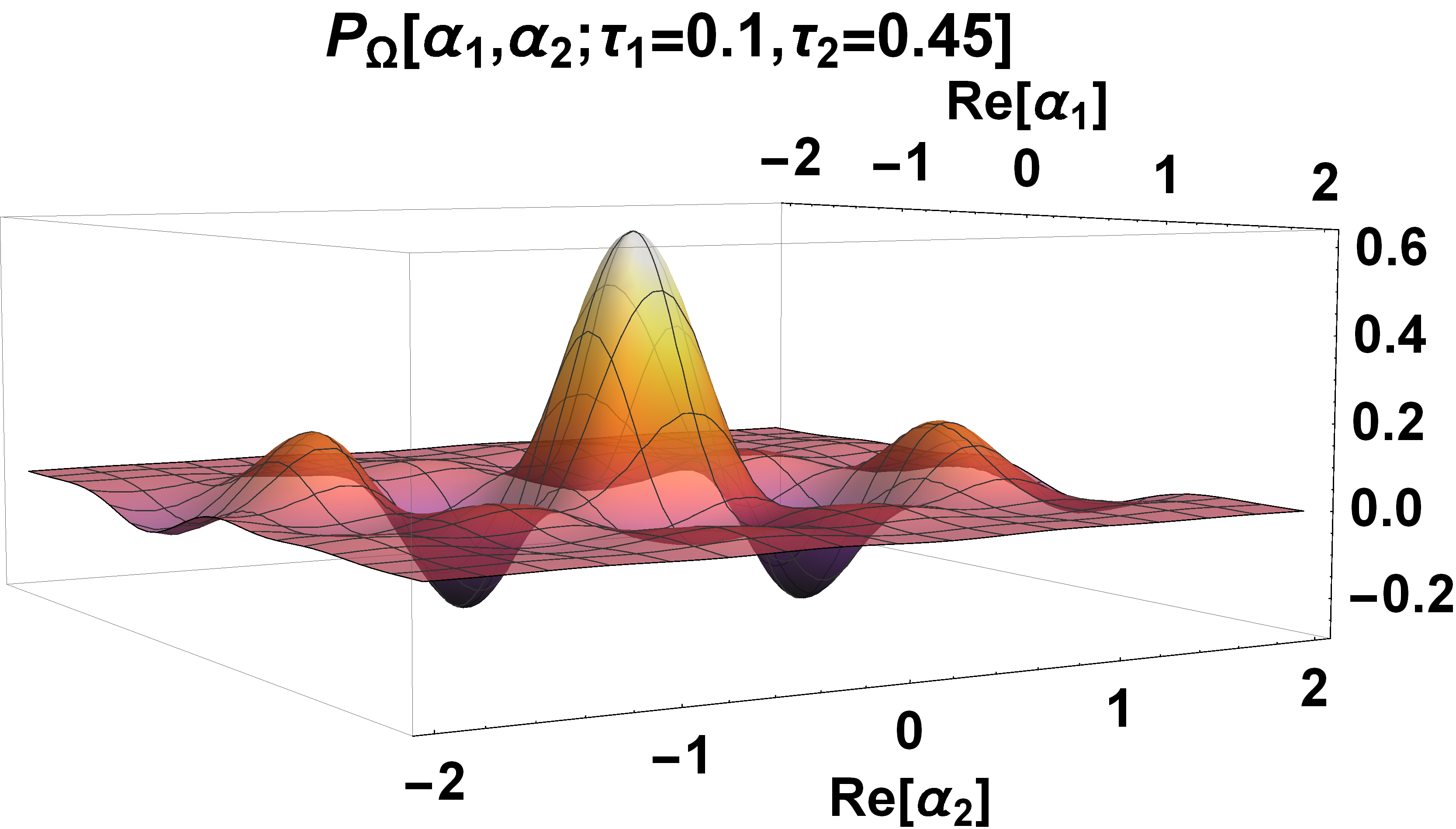}}
	\caption{
		Plot of the two-time regularized functional $P_\Omega(\alpha_1,\alpha_2;\tau_1,\tau_2)$ for $w=2.9$.
		Here, we used $\tau_1=0.1$, $\tau_2=0.45$, and the cross section ${\rm Im} [\alpha_2]={\rm Im} [\alpha_1]=0$. 
		The negativities of the quasiprobability reveal nonclassical normally and time-ordered correlation properties of the considered system.
	}\label{Fig:multi-timeFunctional}
\end{figure}

	The two-time regularized $P$ functional, \eqref{Eq:Pfunctionmulti-timeRegul}, is depicted in Fig. \ref{Fig:multi-timeFunctional}.
	The negativities directly reveal the quantum correlations of the system under study.
	Note that for any time pairings $(\tau_{1},\tau_{2})$ and cross sections other than those used in Fig. \ref{Fig:multi-timeFunctional}, negativities are revealed as well; cf. also Ref. \cite{KSV16}.
	Let us stress that the existence of negativities for a certain filter width is necessary and sufficient for the existence of quantum correlations within the singular $P$ functional.

\section{Trapped-ion dynamics} \label{Sec:TrappedIon}
	
	In the previous section, we discuss multitime effects for a time-dependent parametric oscillator.
	In this case, however, the rather simple commutator $[\hat a(t),\hat a(t')]\propto \hat 1$ for $t\neq t'$ holds [Eq.~\eqref{Eq:MTcommrel}], and, hence, the standard BCH formula is applicable.
	Furthermore, one is able to solve the equations of motion analytically.
	The question arises how such systems shall be examined if the exact dynamics is not explicitly given.
	In this section, we therefore study more general structures of the MTCF and the corresponding dynamics for which the commutators are not central---i.e., they do not commute with the field operators itself, $[[\hat a(t),\hat a(t')],\hat a(t^{(\prime)})]\neq0$.

	Let us first rewrite the MTCF [Eq.~\eqref{Eq:multi-timeCF}] for the two-time case $k=2$ and an input state $\hat \rho_{\rm in}$,
	\begin{eqnarray}
		& & \Phi(\beta_1,\beta_2;t_1,t_2) \\
		= & & \mathrm{Tr} \left[ \hat \rho_{\rm in}(t_1,t_0)  e^{\beta_1 \hat a^\dag} \hat D(\beta_2;t_2,t_1)
		e^{-\beta_1^\ast \hat a} \right] e^{|\beta_2|^2/2}. \nonumber
	\end{eqnarray}
	Here we have used $ \hat \rho_{\rm in}(t_1,t_0)  = \hat U(t_1,t_0) \hat \rho_{\rm in} \hat U(t_1,t_0)^\dag $ and the time-evolved displacement operator
	$\hat D(\beta_2;t_2,t_1)= \hat U(t_2,t_1)^\dag \hat D(\beta_2) \hat U(t_2,t_1)$.
	This form of $\Phi$ reveals a major difficulty:
	$e^{\beta_1 \hat a^\dag}$ and $e^{-\beta_1^\ast \hat a}$ are unbounded operators~\cite{PB15}, and they cannot be rewritten in a simple manner when the commutators are not central.

	Let us therefore consider the limit $t_1\to t_0$ and $t_0\to 0$, i.e., $t_1=t_0=0$.
	As in general $[\hat a(0),\hat a(t_2)] \neq 0$ (even $\not \propto \hat 1$) holds true, this situation differs from the single-time case.
	Let us emphasize that the time-ordering prescription still applies.
	Setting $t_2 \equiv t$, we arrive at
	\begin{eqnarray}
	\label{Eq:MTCFspecialcase}
		& & \Phi(\beta_1,\beta_2;0,t) \\
		= & & \mathrm{Tr} \left[ \hat \rho_{\rm in} e^{\beta_1 \hat a^\dag} \hat D(\beta_2;t)
		e^{-\beta_1^\ast \hat a} \right] e^{|\beta_2|^2/2}. \nonumber
	\end{eqnarray}
	Evaluations of this expression depend on the input state $\hat \rho_{\rm in}$ and on the dynamics under study.
	Here we focus on Fock states as input states, i.e., $\hat \rho_{\rm in}=|p\rangle \langle p|$.
	This yields
	\begin{eqnarray}
	\label{Eq:MTCFspecialcase2}
		& & \Phi(\beta_1,\beta_2;0,t) \\
		= & & \sum_{m,n=0}^p\frac{\beta_1^n(-\beta_1^\ast)^m} {m!n!}\frac{p!\langle p-n| \hat D(\beta_2;t) |p-m\rangle }{\sqrt{(p-n)!(p-m)!}}  e^{|\beta_2|^2/2} , \nonumber
	\end{eqnarray}
	which is obtained via expanding the exponential functions in power series and using the standard actions of $\hat a$ ($\hat a^\dag$) on the Fock states $|p\rangle$ ($\langle p|$).
	
	First, we study Eq.~\eqref{Eq:MTCFspecialcase2} for $p=0$, i.e., $\hat \rho_{\rm in}=|{\rm vac}\rangle \langle {\rm vac}|$.
	We obtain
	\begin{equation}
		\Phi_{\rm vac} (\beta_1,\beta_2;0,t) = \langle0  | \hat D(\beta_2;t) | 0 \rangle e^{|\beta_2|^2/2},
	\end{equation}
	which is always bound by an inverse Gaussian factor.
	Furthermore, $\Phi_{\rm vac} (\beta_1,\beta_2;0,t)$ equals a single-time characteristic function, and the corresponding $P$~functional [Eq.~\eqref{Eq:FourierFunctional}] will attain for any dynamics the form
	\begin{equation}
		P_{\rm vac} [\alpha_1,\alpha_2;0,t]= P_{\rm vac}[\alpha_2;t] \delta(\alpha_1).
	\end{equation}
	As the delta distribution is a nonnegative distribution, the two-time $P$~functional describes a nonclassical system if and only if the single-time $P$~function fails to be a classical probability distribution.
	In this scenario, there are no genuine temporal correlations.
	However, the situation is different for other input states, which can be observed for our second example, $p=1$.
	 Because $e^{-\beta_1^\ast \hat a}|1  \rangle= |1  \rangle - \beta_1^\ast |0 \rangle$, we get
	\begin{eqnarray}
		    & & \Phi_{1} (\beta_1,\beta_2;0,t) \nonumber \\
		    =   & & \left[ \langle 1 |  + \langle 0|  \beta_1 \right]   \hat D(\beta_2;t) \left[ |1  \rangle - \beta_1^\ast |0 \rangle \right]  e^{|\beta_2|^2/2},
	\end{eqnarray}
	and, therefore, additional terms due to the inclusion of a second point in time.
	 This holds even if we set the first time to be $0$.
	Note, a similar behavior can be observed for any $p\geq 1$.

	For clarification, let us consider a realistic interaction Hamiltonian~\cite{WV97},
	\begin{equation}
		\label{Eq:trappedion}
		\hat H_3= \hbar \varepsilon \hat f_3 (\hat a^\dag \hat a;\eta) (i \eta \hat a)^3 + {\rm H.c.},
	\end{equation}	
	which results in a time evolution obeying a noncentral commutator algebra and whose time evolution is solved numerically.
	It describes a nonlinear vibrational dynamics of a laser-driven trapped ion.
	Therein, $\varepsilon$ is the effective two-photon coupling strength, and $\eta$ is the Lamb-Dicke parameter.
	The nonlinear operator function $\hat f_3 (\hat a^\dag \hat a;\eta) $ of the vibrational number operator $\hat a^\dag \hat a$ accounts for the recoil effects due to absorption and emission of laser photons by the trapped atom.
	It reads~\cite{WV97}
	\begin{equation}
		\hat f_3 (\hat a^\dag \hat a;\eta)  = e^{-\eta^2/2}\sum_{l=0}^\infty (-1)^l \frac{\eta^{2l}}{l! (l+3)!} \hat a^{\dag l} \hat a^l.
	\end{equation}
	The action on Fock states of the ion's center-of-mass motion yields
	\begin{equation}
		\hat f_3(\hat a^\dag \hat a;\eta) |n \rangle = e^{-\eta^2/2} \frac{n!}{(n+3)!} L_n^{(3)}(\eta^2) | n \rangle \equiv f_3(n;\eta) |n \rangle,
	\end{equation}
	with $L_n^{(k)}(x)$ being the generalized Laguerre polynomials.
	Using the completeness relation of the Fock states, $\sum_{n=0}^\infty |n\rangle \langle n|=\hat 1$, the Hamiltonian, \eqref{Eq:trappedion}, can be written in the Fock basis as
	\begin{equation}
		\hat H_3
		=  i \hbar \varepsilon \eta^3 \sum_{n=0}^\infty g_3(n;\eta) |n+3\rangle \langle n| - g_3(n;\eta) |n\rangle \langle n+3|, 
	\end{equation}
	with $g_3(n;\eta)=f_3(n;\eta)\sqrt{(n+1)(n+2)(n+3)}$.
	
	The time evolution can be numerically solved via evaluating the time evolution operator 
	\begin{equation}
		\hat U(t,t_0)=\exp \left[ - \frac{i}{\hbar} (t-t_0) \hat H_3 \right],
	\end{equation}
	in matrix representation, where a sufficiently high cutoff of the Fock space has to be chosen.
	Using the MTCF in Eq.~\eqref{Eq:MTCFspecialcase}, we can investigate the difference between the squared modulus of the MTCF and the inverse Gaussian bound,
	\begin{equation}
		\label{Eq:fom}
		\Delta \Phi(\beta_1,\beta_2;0,\tau) := | \Phi(\beta_1,\beta_2;0,\tau)|^2 - e^{|\beta_1|^2+|\beta_2|^2},
	\end{equation}
	with the dimensionless system time $\tau \equiv  \varepsilon t$.
	If this function $\Delta \Phi$ exceeds $0$, the MTCF is---due to temporal correlations---differently bounded compared to the two-mode single-time characteristic function.
	This means that the temporal correlations increase the divergences of the $P$~functional to an extent which cannot occur for any two-mode correlations at equal time; cf. Eq. \eqref{Eq:multi-mode}.

\begin{figure}[tb]
	\centering
	{\includegraphics*[width=8cm]{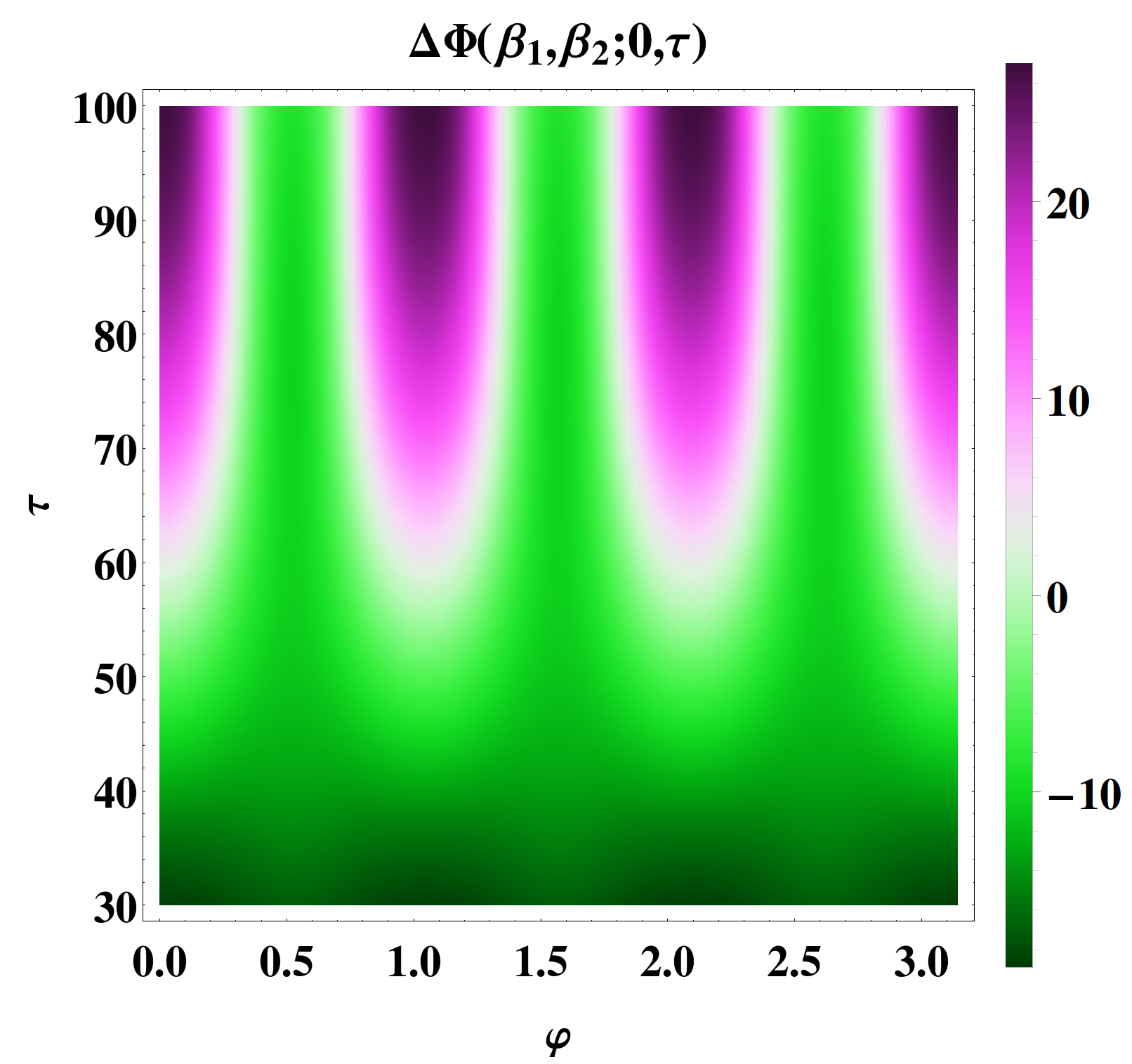}}
	\caption{(Color online)
		Plot of $\Delta \Phi$ as defined in Eq.~\eqref{Eq:fom} for fixed $|\beta_1|=|\beta_2|=1.3$ and $\varphi_1=\varphi_2\equiv \varphi$, where $\beta_j=|\beta_j| e^{i \varphi_j}$, $j=1,2$.
		We vary the common phase $\varphi$ and the dimensionless time $\tau=\varepsilon t$.
		As $\Delta \Phi$ clearly exceeds the value of $0$ (pink areas), we confirm that the MTCF is differently bounded compared to the two-mode single-time case.
	}\label{Fig:mtcftrapped}
\end{figure}

	A visualization of \eqref{Eq:fom} is given in Fig.~\ref{Fig:mtcftrapped} for a particular choice of the parameters.
	We used $p=3$ ($\hat \rho_{\rm in}=|3\rangle \langle 3|$) and numerically evaluated $\Delta \Phi$ in a 200-dimensional Fock space to ensure approximation errors of the order of those of the numerical arithmetic.
	As for several parameters $\Delta \Phi >0$ holds, the temporal correlations obviously exceed the slope one can maximally expect for an equal-time two-mode characteristic function.
	As discussed earlier, the strength of the excess depends on the chosen input state.
	This deviation from the inverse Gaussian bound of the MTCF could be even stronger for other dynamical systems.
	However, our filter approach, introduced in Sec. \ref{Subsec:UMR}, is suitable for regularization of the $P$~functional for any dynamics.
	Based on the strongly nonlinear trapped-ion interaction Hamiltonian and the discussion of the impact of the input state, we have demonstrated our approach's requirement for regularization of multitime $P$~functionals.

\section{Sampling of the $P$ functional}\label{Sec:Measurement}

\begin{figure*}[tb]
	\centering
	{\includegraphics*[width=16cm]{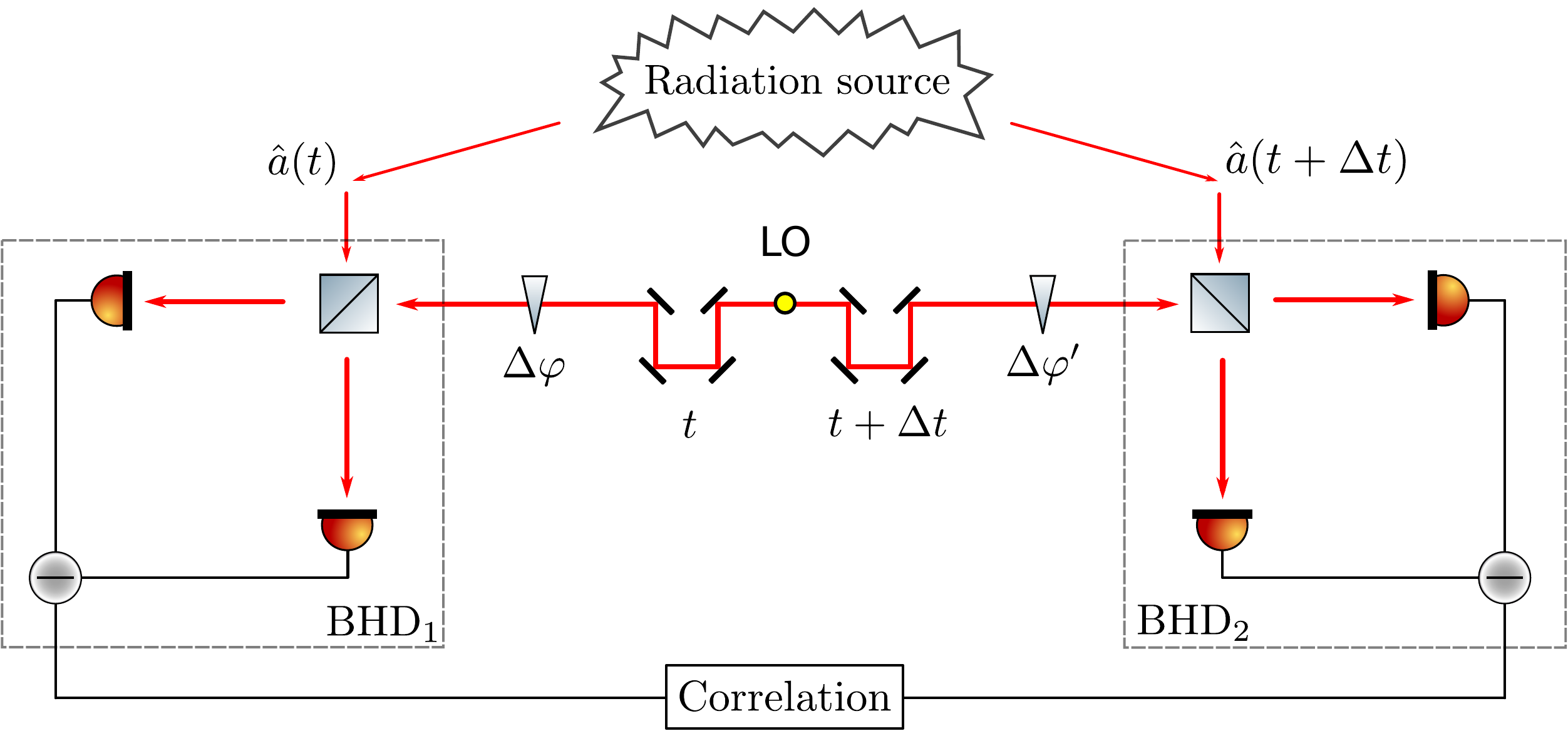}}
	\caption{
		Experimental scheme to directly measure two-time quantum correlations in terms of two-time quasiprobabilities.
		The scheme consists of two balanced homodyne detection (BHD) setups whose difference signals are additionally correlated.
		The creation operators $\hat a(t)$ and $\hat a(t+\Delta t)$ label the different travel times of the field.
		The resulting correlated difference statistics of the detector events can be directly related to $P_\Omega[\alpha_1,\alpha_2;\tau_1,\tau_2]$, 
		where the phases for the local oscillator (LO) of each BHD setup are controlled continuously through the phase shifters $\Delta \varphi$ and $\Delta \varphi'$.
		The temporal matching of the LO to the signal fields is ensured through a proper path-length control.
	}\label{Fig:Setup}
\end{figure*}

	The first experimental reconstruction of a phase-insensitive and single-time filtered $P_\Omega$~function with negativities was performed for single-photon-added 
	thermal states \cite{KVBZ11}---based on the measurement of quadratures in balanced homodyne detection.
	A direct sampling of a single-time $P_\Omega$~function for a squeezed state was then performed \cite{KVHS11}---including the formulation of suitable pattern functions for discrete phase measurements.
	More recently, a method was presented to sample the filtered $P$ function via continuous-in-phase measurement \cite{ASVKMH15}.
	In the following  we study an optical measurement scheme for reconstructing the filtered $P$ functional that allows one to apply our technique in experiments.
	Additionally note that a corresponding measurement technique can also be provided for the motional quantum state of a trapped ion, which can be directly based on the motional-state reconstruction as proposed in Ref.~\cite{Wa-VO95}.

	The setup of our scheme is shown in Fig. \ref{Fig:Setup}; it correlates the radiation field at two times.
	According to the quantum theoretical model for photodetection \cite{VW06,KK64,G65}, the joint probability of four detectors is
	\begin{equation}
		\label{Eq:photodetection}
		P_{n_1,n_2,n_3,n_4}=\left\langle \begin{smallmatrix}\circ \\ \circ\end{smallmatrix} \prod_{i=1}^{4}  \frac{[\eta_i \hat n_i(t_i)]^{n_i}}{n_i!} e^{-\eta_i \hat n_i(t_i)}   \begin{smallmatrix}\circ \\ \circ\end{smallmatrix} \right\rangle,
	\end{equation}
        where $n_i$ denotes the number of photons at the $i$th detector, $\eta_i$ is the detector efficiency, and $\hat n_i(t_i)$ is the photon number operator at time $t_i$ in the corresponding detector path.
	As we correlate two points in time in our setup, we set $t_1=t_2\equiv t$ and $t_3=t_4\equiv t+\Delta t$.
	The phases of the local oscillator modes can be controlled separately by the phase shifters $\Delta \varphi$ and $\Delta \varphi'$.
	Note that the layout is scalable and could be further extended to more points in time.

        In close analogy to the procedure in Ref. \cite{VG93}, the correlated difference statistics is found to be 
	\begin{eqnarray}
		\label{Eq:measureddiffstatistics}
		& &p_{t,t+\Delta t}(v,v' ;\varphi,\varphi')
		\\\nonumber
		&=&\frac{1}{2 \pi R^2 \sqrt{\eta \eta'}}
		\\& & \times \Big\langle \begin{smallmatrix}\circ \\ \circ\end{smallmatrix} 
		\exp\left\{ {-}\frac{[v {-} \eta R  \hat x(\varphi-\pi/2;t) ]^2}{2 \eta R^2} \right\} \nonumber \\
		& & \times \exp\left\{  {-}\frac{[v'  {-}  \eta' R  \hat x(\varphi'-\pi/2;t{+}\Delta t) ]^2}{2 \eta' R^2} \right\} \begin{smallmatrix}\circ \\ \circ\end{smallmatrix} \Big\rangle,\nonumber
	\end{eqnarray}
	with the difference events $n_1-n_2 = v$ and $n_3-n_4 = v' $, the phases $\varphi$ and $\varphi'$, the common amplitude $R$ of the two local oscillators, and the quadrature operator
	\begin{eqnarray}
		\hat x(\varphi-\pi/2;t)  &=& \hat a(t) e^{-i \varphi}+\hat a(t)^\dag e^ {i\varphi} . 
	\end{eqnarray}
	To arrive at Eq. \eqref{Eq:measureddiffstatistics}, we replaced $n_i$ with continuous variables, which can be done in the strong local oscillator limit and for a sufficiently large number of events.
	The distribution, \eqref{Eq:measureddiffstatistics}, is obviously the quantum expectation value of the time- and normal-ordered product of two Gaussian distributions of the quadratures at $t$ and $t+\Delta t$.
	Via the two-dimensional Fourier transform of the measured difference statistics,
	\begin{eqnarray}
		& & F_{t,t+\Delta t}(y,y',\varphi,\varphi)\\
		= & & \int {d}v \int {d}v' e^{i v y } e^{i v'  y'} p_{t,t+\Delta t}(v ,v' ;\varphi,\varphi'),  \nonumber 
	\end{eqnarray}
	one gets
	\begin{eqnarray}
		& &F_{t,t+\Delta t}(y,y',\varphi,\varphi')e^{y^2 R^2 \eta/2+y'^2R^2 \eta'/2}  \\
		=& & \Phi(y\eta R e^{i \varphi},y' \eta' R e^{i\varphi'};t,t+\Delta t). \nonumber
	\end{eqnarray}
	Here $\Phi$ is the desired two-time characteristic function of the $P$ functional.
	Hence, by adjusting the parameters and properly scaling the complex numbers $y$ and $y'$, $\Phi$ can be directly sampled with our setup.
	If we use the representations $\beta_1=b_1 e^{i \varphi}$ and $\beta_2= b_2 e^{i\varphi'}$ and identify $y\eta R \equiv b_1$ and $y'\eta'R\equiv b_2$, we can rewrite the previous results as
	\begin{eqnarray}
		\label{Eq:expTTcharfct}
		& & \Phi(b_1 e^{i \varphi},b_2 e^{i\varphi'};t,t+\Delta t)   \\ 
		= & &  \exp \left [ \frac{b_1^2}{2 \eta}+\frac{b_2^2}{2 \eta'} \right]  \int {d}v \int {d}v'  \, p_{t,t+\Delta t}(v,v' ;\varphi,\varphi' )   \nonumber \\
		& & {\times}   \exp \left[ \frac{ib_1}{\eta R} v  {+} \frac{ib_2}{\eta' R} v'   \right]. \nonumber
	\end{eqnarray}
	
	The regularized $P$ functional can be reconstructed in the following way.
	First, we recall that [cf. Eq. \eqref{Eq:multi-timefiltering2} for $k=2$]
	\begin{eqnarray}
	& & P_\Omega [\alpha_1,\alpha_2;t,t+\Delta t]\\
	= & & \int\! \frac{{d}^2\beta_1}{\pi^2} e^{\beta_1^{\ast}  \alpha_1{-}\beta_1   \alpha_1^{ \ast}}  \int\! \frac{{d}^2\beta_2}{\pi^2} e^{\beta_2^{\ast}  \alpha_2{-}\beta_2   \alpha_2^{ \ast}}   \nonumber \\
		& & \times\Phi(\beta_1,\beta_2;t,t+\Delta t) \Omega_w(\beta_1) \Omega_w(\beta_2) . \nonumber 
	\end{eqnarray}
	Since we have used a product filter [cf. Eq. \eqref{Eq:productfilter}], we can simply insert Eq. \eqref{Eq:expTTcharfct} and, following the procedure in Ref. \cite{KVHS11}, rewrite the previous formula as
	\begin{eqnarray}
		\label{Eq:prepattern}
		& & P_\Omega [\alpha_1,\alpha_2;t,t+\Delta t]
		\\&=&\nonumber
		\int {d}v \int {d}v'  \int_{0}^{\pi} {d}\varphi \int_{0}^{\pi} {d}\varphi'  \frac{p_{t,t+\Delta t}(v,v' ;\varphi,\varphi' )}{\pi^2}
		\\& &\times\nonumber
		f_\Omega(v,\varphi;\alpha_1,w) f_\Omega(v' ,\varphi' ;\alpha_2,w).
	\end{eqnarray}
	Here, the so-called patten function $f_\Omega$ is
	\begin{eqnarray}
		\label{Eq:patternfunction}
		f_\Omega(z,\varphi;\alpha_i,w) =& &\int  {d}b_i \frac{b_i}{\pi}  \Omega_w(b_i)  \exp\Big[{ \frac{i  z b_i}{\eta_{z} r} }+{\frac{b_i^2}{2 \eta_{z}}}\Big]\\
		& & \times   \exp[{2i b_i |\alpha_i| \sin[\varphi_{\alpha_i} -  \varphi - \pi/2]}], \nonumber
	\end{eqnarray}
	with $z \in \{ v,v' \}$, $\eta_{v} \equiv \eta$, and $\eta_{v' } \equiv \eta'$.
	Finally, the regularized $P$ functional can be sampled from $M$ measured quadrature data points $(v_j,\varphi_{j},v_j',\varphi_{j}')_{j=1}^M$ in the two channels via its empirical estimate
	\begin{eqnarray}
		& & P_\Omega [\alpha_1,\alpha_2;t,t+\Delta t]
		\\& \approx &\nonumber
		\frac{1}{M} \sum_{j=1}^{M}  f_\Omega(v_j,\varphi_j;\alpha_{1},w) f_\Omega(v_j' ,\varphi_j';\alpha_{2},w),
	\end{eqnarray}
	where the time dependences are included in the set of data, $v\equiv v(t)$ and $v'\equiv v(t+\Delta t)$. 
	For convenience, we used a radial symmetric filter, such that the filter functions $\Omega_w$ depend only on the $b_i$ and not on the phases $\varphi$ and $\varphi'$ [cf. Eqs. \eqref{Eq:prepattern} and \eqref{Eq:patternfunction}].

	Hence, we have formulated the theory of our proposed measurement scheme in Fig. \ref{Fig:Setup} that renders it possible to directly obtain the two-time regularized $P$ functional via the sampling of measured (correlated) quadrature data using pattern functions.
	We may stress again that this approach is scalable to an arbitrary number of points in time by employing multiple BHDs.
	In addition, some remarks concerning the sampling error estimation can be found in Ref. \cite{ASVKMH15}.

\section{Summary and Conclusions}\label{Sec:Conclusions}

	In summary, we have derived a method for visualizing general multitime quantum correlations in terms of regular phase-space quasiprobabilities.
	For this purpose, we significantly generalized the approach of filtering the Glauber-Sudarshan $P$ function for a single time to a regularization of the $P$ functional that correlates an arbitrary number of points in time.
	This generalizes the commonly accepted definition of nonclassicality by Titulaer and Glauber to capture multitime quantum phenomena in terms of regular and accessible phase-space distributions.
	Hence, our formulation of multitime and regular nonclassicality quasiprobabilities is not restricted to particular correlation functions or observables.
	Additionally, we have proved that our method is applicable to arbitrarily complex evolutions of light fields.
	In our general approach, nonclassical correlations, if present, are directly visualized by negativities of this regular quasiprobability density for properly chosen filter widths.
	Beyond previously studied methods, our treatment enables us to visualize general quantum correlations of radiation fields, including quantum entanglement as a subset.

	We applied this technique to characterization of the temporal quantum features of a parametric oscillator with a frequency mismatch.
	We studied the impact of the time-dependent Hamiltonian on the dynamical properties of this system, including the multitime correlations.
	In particular, we uncovered quantum correlations via a negative and regular two-time quasiprobability description of this process.
	We also studied a strongly nonlinear dynamics of the motional quantum state of a laser-driven trapped ion. 
	In this case, nontrivial time-dependent commutator rules become important, leading to singularities of the $P$ functional much stronger than those occurring for equal-time two-mode correlations.
	Our technique regularizes these unexpectedly strong singularities.
	Eventually, we presented an experimental setup---consisting of two correlated balanced homodyne detection layouts.
	Based on the derived pattern functions, this allows one to directly sample the regularized quasiprobabilities in experiments.
	Altogether, this yields a powerful tool for the characterization of general, time-dependent quantum correlations in phase space.

\begin{acknowledgments}
	We thank Regina Kruse for helpful comments.
	This work was supported by the Deutsche Forschungs\-gemeinschaft through SFB~652, project B12.
	J.S. and W.V. acknowledge funding from the European Union's Horizon 2020 research and innovation program under Grant Agreement No. 665148.
\end{acknowledgments}


\begin{thebibliography}{100}
\bibitem{comment}
		A function with a compact support decays more rapidly than any polynomial.
		Thus, due to Sobolev's lemma, all orders of derivatives of its  Fourier transform exist; see, e.g., Refs. \cite{S16} and \cite{ASV13} and references therein.
\end{thebibliography}
\end{document}